\documentclass{sig-alternate-05-2015}

\usepackage{balance}
\usepackage[font={small}]{caption}
\usepackage{color}
\usepackage{makecell}
\usepackage{amsmath}
\usepackage{multirow}
\usepackage{graphicx}
\usepackage{hhline}
\usepackage[breaklinks]{hyperref}
\usepackage{microtype}
\usepackage{wrapfig}
\usepackage[subtle]{savetrees}
\usepackage{balance}
\usepackage[T1]{fontenc} 
\usepackage{tgtermes}
\usepackage{paralist}

\usepackage{enumitem}

\setcopyright{none}

\makeatletter
\def\@copyrightspace{\relax}
\makeatother

\usepackage{fancyhdr}
\fancypagestyle{plain}{%
   \fancyhf{} %
   \fancyfoot[L]{ACM SIGCOMM Computer Communication Review }%
   \fancyfoot[R]{Volume 47 Issue 3, July 2017}%
}
\def\refname{}

\begin{document}

\title{Geohyperbolic Routing and Addressing Schemes}

\numberofauthors{4}
\author{
\begin{tabular*}{0.6\textwidth}%
{@{\extracolsep{\fill}}cc}
Ivan Voitalov & Rodrigo Aldecoa\\
\affaddr{Northeastern University, USA} & \affaddr{Northeastern University, USA}\\ 
\email{i.voitalov@northeastern.edu} & \email{raldecoa@northeastern.edu}
\end{tabular*}  \vspace{3mm} \\
\begin{tabular*}{0.6\textwidth}%
{@{\extracolsep{\fill}}cc}
Lan Wang & Dmitri Krioukov\\
       \affaddr{University of Memphis, USA} & \affaddr{Northeastern University, USA}\\
       \email{lanwang@memphis.edu} & \email{dima@northeastern.edu}
\end{tabular*}\\
\begin{tabular}{c}
\end{tabular}\\
}

\maketitle

\begin{abstract}
The key requirement to routing in any telecommunication network, and especially in Internet-of-Things (IoT) networks, is scalability. Routing must route packets between any source and destination in the network without incurring unmanageable routing overhead that grows quickly with increasing network size and dynamics. Here we present an addressing scheme and a coupled network topology design scheme that guarantee essentially optimal routing scalability. The FIB sizes are as small as they can be, equal to the number of adjacencies a node has, while the routing control overhead is minimized as nearly zero routing control messages are exchanged even upon catastrophic failures in the network. The key new ingredient is the addressing scheme, which is purely local, based only on geographic coordinates of nodes and a centrality measure, and does not require any sophisticated non-local computations or global network topology knowledge for network embedding. The price paid for these benefits is that network topology cannot be arbitrary but should follow a specific design, resulting in Internet-like topologies. The proposed schemes can be most easily deployed in overlay networks, and also in other network deployments, where geolocation information is available, and where network topology can grow following the design specifications.
\end{abstract}




\section{Introduction}
\label{sec:intro}

The key motivation for this work comes from designing maximally scalable routing in overlay networks. Overlay networks are widely used in testing new network architectures and applications as they allow new software and functionality to be deployed without costly upgrades to existing infrastructure. For instance, most wide-area testbeds, e.g., PlanetLab~\cite{chun2003planetlab}, GENI~\cite{berman2014geni}, or NDN testbed~\cite{ndntestbedurl}, are overlays as testbed nodes are usually geographically dispersed without direct links among them.
A primary goal of any telecommunication network, including overlays, is to efficiently route data packets between any pair of nodes, without incurring unscalable routing overhead.
Two basic routing scalability metrics are \emph{FIB sizes} and \emph{message overhead}, a number of routing protocol messages exchanged upon topology changes. In existing Internet interdomain routing (BGP), for instance, the former metric grows linearly with the number of destination prefixes in the Internet~\cite{huston2017bgp}, while the latter is unbounded due to possibility of persistent routing oscillations~\cite{varadhan2000persistent}. Routing with such scaling properties can clearly NOT meet the demands of fast-evolving data-centric applications and architectures, such as the Internet of Things (IoT) in general, and Named Data Networking (NDN) in particular, as they require immense, essentially unbounded and highly dynamic addressing/destination spaces~\cite{dhumane2016routing}, calling for radical rethinking of how routing should be done in future networks.

\emph{Greedy geometric routing (GGR)}~\cite{papadimitriou2005conjecture} provides a basis for maximally scalable routing. In GGR, all nodes in the network are mapped to a geometric space, i.e., are assigned geometric \emph{addresses}, or coordinates, in this space. Given a source-destination pair of nodes, and the destination node address in the packet, GGR forwards the packet from the current node to its neighbor that is closest to the destination in the space. The packet is forwarded this way until it reaches either the destination node, signifying \emph{routing success}, or a previously visited node, signifying \emph{routing failure} and \emph{local minimum}: a packet circulates within a loop of nodes that do not have any neighbors closer to the destination than themselves.

GGR guarantees the \emph{smallest possible FIB sizes}, since each node has to store only the addresses of the adjacent nodes, and no per-destination information. The percentage of node pairs that can successfully communicate via GGR is called the \emph{success ratio (SR)}. If the \emph{combination of network topology and addressing} is such that GGR's SR is $100\%$ not only in a given static topology, but also in dynamic growing (new nodes/links) or damaged (failing nodes/links) topologies, then no exchange of routing protocol messages is needed, \emph{reducing the message overhead to zero}. However, if SR degrades quickly in dynamic topologies, an \emph{auxiliary routing protocol} is needed to repair broken routes by finding alternative ones. Such protocols would re-introduce the message overhead, which would be the larger, the smaller the SR. Therefore, a satisfactory combination of addressing and topology must be such that GGR's SR stays as close as possible to $100\%$ even if the network continues to grow indefinitely, and even if it undergoes severe connectivity disruptions.

In~\cite{boguna2009navigability,papadopoulos2010greedy} one combination of network topology and addressing satisfying these requirements was identified. In that combination, the topology must be Internet-like. That is, it must have the \emph{power-law} distribution $P(k)\sim k^{-\gamma}$ of node degrees $k$ with exponent $\gamma=2$, and it must also have as many triangular subgraphs as possible, i.e., strongest possible \emph{clustering}. The addressing must be \emph{hyperbolic}, meaning that the coordinates of nodes are the coordinates of points sprinkled quasi-uniformly at random over the hyperbolic plane~\cite{krioukov2010hyperbolic}. Since the Internet Autonomous System (AS) topology is Internet-like, it can be and was embedded into the hyperbolic plane by finding the coordinates of all AS nodes~\cite{boguna2010sustaining,papadopoulos2015network,papadopoulos2015neighbors}. Yet three key difficulties with this approach are that: 1)~it requires either full or partial knowledge of the given network topology, 2)~the embedding coordinate computations are not entirely straightforward, and 3)~they do not account for network delays, the minimization of which is yet another efficiency goal of routing.

Here we present a coupled combination of the \emph{network design scheme}, Section~\ref{sec:design}, and \emph{addressing scheme}, Section~\ref{sec:models}, that resolve all the issues above, while maintaining the remarkable robustness of hyperbolic GGR. The addressing scheme is \emph{geohyperbolic~(GH)}. It does not require any sophisticated computations. The addresses are determined by a simple formula that relies only on the geographic coordinates of nodes and a measure of centrality. Given these coordinates, one first maps them to a node's centrality score. The centrality scores can be anything that geographic coordinates uniquely determine. Here for concreteness we consider population density as the centrality score. Given the geographic coordinates and centrality scores, the latter determined by the former, the node is then mapped to a unique location in the \emph{geohyperbolic space}, equivalent to the three-dimensional hyperbolic space~$\mathbb{H}^3$. The network design scheme is trivial. As a new node joins the network, it must connect to a small fixed number of nodes already in the network that are closest to the new node in~$\mathbb{H}^3$.

We investigate, in Section~\ref{sec:simulations}, the performance of GH GGR in large synthetic networks representing a worldwide overlay network deployment, and in the existing NDN testbed of 33 nodes built according to the proposed network design scheme. We find that GH GGR's SR is nearly $100\%$, degrading only to $97\%$ even upon catastrophic damages to the network ($20\%$ links or nodes failing), and that these SR's values are essentially independent of the size of synthetic networks. GH GGR's delays are slightly higher than in the delay-optimal purely geographic (GEO) scheme, whose SR degrades quickly to unacceptable values as the network grows or gets damaged. However, a minor improvement of vanilla GH, the \emph{regionalized (R) GH} scheme, leads to nearly optimal delays, while maintaining remarkable SR robustness.

We discuss possible extensions of (R)GH to other network scenarios in Section~\ref{sec:discussion}, related work in Section~\ref{sec:related}, and conclude with final remarks in Section~\ref{sec:conclusion}.

\section{Scheme Specifications}

\subsection{Routing and network design schemes}
\label{sec:design}

The routing and network design schemes are \emph{the same} in \emph{all} the addressing schemes considered below. \emph{The routing scheme} is pure GGR~\cite{papadimitriou2005conjecture}, described in the Introduction.

\emph{The network design scheme} prescribes any new node joining the network to connect to a fixed number $m$ of existing nodes that are closest to the new node in the geometric space. This space and the coordinates of existing and new nodes in it are determined by an addressing scheme, so that this procedure is well defined. Parameter $m$ is the only parameter of the design scheme, defining the average degree $\bar{k}=2m$ in the resulting topology.

\subsection{Addressing schemes}\label{sec:models}

\textbf{Geographic scheme (GEO).} The GEO scheme is used in geographic routing~\cite{karp2000greedy}. We refer to it here for baseline comparisons in the next section. It is the simplest possible scheme in geolocation-based routing: addresses of nodes are their geographic coordinates. Since connections follow geography, and RTT delays in the Internet are linearly correlated with geographic distances~\cite{huffaker2002distance,kaune2009modelling,landa2013large}, the scheme is essentially delay-optimal. Yet we will see that both this delay optimality and, more importantly, SR are extremely fragile w.r.t. to both network growth and failures.

In the scheme, the address of node $i$ is its latitude and longitude mapped to angular coordinates $(\theta_i, \phi_i)$ on two-dimensional sphere $\mathbb{S}^2$. The greater circle distance between a pair of nodes $i$ and $j$, used by the network design scheme and GGR, is
\setlength{\abovedisplayskip}{1pt}
\setlength{\belowdisplayskip}{1pt}
\begin{gather}
d_{ij} = R_{E} \Delta \theta_{ij}, \label{eq:geodist} \\
\Delta \theta_{ij} = \cos^{-1}\left(\cos{\theta_i}\cos{\theta_j} + \sin{\theta_i}\sin{\theta_j}\cos{(\phi_i - \phi_j)}\right), \label{eq:angdist}
\end{gather}
where $R_{E} = 6371$km is the radius of the Earth, and $\Delta \theta_{ij}$ is the central angle between nodes $i$ and $j$ on $\mathbb{S}^2$.

\textbf{Geohyperbolic scheme (GH).} The GH scheme can be thought of as a ``hyperbolization'' of the GEO scheme. This hyperbolization is achieved by assigning to each node $i$ an additional third radial coordinate $r_i$, while the other two angular coordinates $(\theta_i, \phi_i)$ are still the geographic coordinates. The radial coordinate is treated as the radial coordinate in the spherical coordinate system in the three-dimensional hyperbolic space~$\mathbb{H}^3$. The distances in $\mathbb{H}^3$, used by the design scheme and GGR, between a pair of nodes $i$ and $j$ with coordinates $(r_i,\theta_i,\phi_i)$ and $(r_j,\theta_j,\phi_j)$ is
\setlength{\abovedisplayskip}{2pt}
\setlength{\belowdisplayskip}{2pt}
\begin{equation}
\label{eq:hypdist}
d_{ij}  = \cosh^{-1}\left(\cosh{ r_i}\cosh{ r_j} - \sinh{ r_i}\sinh{ r_j}\cos{\Delta \theta_{ij}}\right),
\end{equation}
where $\Delta \theta_{ij}$ is given by Eq.~(\ref{eq:angdist}). If the distribution of radial coordinates follows the exponential distribution $P(r)\sim e^{\alpha r}$ with exponent $\alpha=1$, then the network design scheme in Section~\ref{sec:design} guarantees that the degree distribution in the resulting network topology follows the power law $P(k)\sim k^{-\gamma}$ with exponent $\gamma=2$, while clustering is strongest possible~\cite{Papadopoulos2012}, thus resulting in a maximally efficient combination of network topology and addressing discussed in the Introduction. The key component of the scheme is thus a \emph{centrality scheme} determining $r_i$ based on  $(\theta_i, \phi_i)$, discussed next.

\textbf{Centrality scheme.} The centrality scheme assigns a radial coordinate $r_i$ in $\mathbb{H}^3$ to \emph{any} geographic location $(\theta_i, \phi_i)$ as follows. Suppose a network is to be deployed over a certain geographic area of interest. It can be the whole Earth surface, a continent, or a country. This area is first tessellated into a fixed number $Q$ of zones $z=1,2,\ldots,Q$, and then a fixed centrality score $s_z\geq1$ is assigned to each zone $z$. The scores can be anything, yet we note that the higher the score, the larger the number of GGR's paths will be routed via nodes with this score, explaining why we call it ``centrality score''. Therefore the general recommendation is to assign higher scores to zones with higher expected centrality in the network. In the next section, we consider the simplest option satisfying these requirements: $s_z$'s are populations of cities $z$, which are distributed as the power law with $\gamma=2$~\cite{gabaix1999zipf}.

The centrality score $s_z$ of each zone $z$ is then mapped to $z$'s \emph{centrality rank} $t_z$, which is $z$'s rank in the list of zones sorted in the \emph{decreasing order of their centrality scores}. Finally, any geolocation $(\theta_i, \phi_i)$ that falls within zone $z$ in the tessellation gets radial coordinate $r_i=r_z$ given by
\vspace{-2mm}
\begin{equation}\label{eq:centrality}
r_z = \ln\left(\xi + t_z\right),
\end{equation}
where $\xi \geq 0$ is the only parameter of the scheme defining the smallest $r_z$, $r_{z,\min}=\ln(\xi+1)$. The larger the $\xi$, the less important the radial coordinates in the hyperbolic distance~(\ref{eq:hypdist}), and the more important the angular geographic distance. Varying $\xi$ thus tunes the relative importance of geography versus centrality-based hyperbolicity, controlling the preference of new nodes to connect to either more geographically closer or more central existing nodes.
Since the radial coordinates $r$ are set equal to the logarithm of node ranks, then the distribution of these coordinates follows the exponential distribution $P(r)\sim e^{\alpha r}$ with exponent $\alpha=1$, so that the degree distribution in the resulting network is the power law $P(k)\sim k^{-\gamma}$ with $\gamma=2$~\cite{Papadopoulos2012}, unless $\xi$ is so large that GH becomes effectively GEO.

\textbf{Regionalized geohyperbolic scheme (RGH).} If the GH scheme is deployed over geographically vast areas, it may result in suboptimal delays. Indeed, consider packet forwarding from Boston to Los Angeles, for instance, and suppose that the most central node is located in Shanghai, e.g., because of its largest population. If the Shanghai zone is closest to the origin of the $\mathbb{H}^3$ space, i.e., if its radial coordinate is smallest, a majority of GGR paths will be routed thought it due to its proximity to a majority of geodesic paths in $\mathbb{H}^3$~\cite{krioukov2010hyperbolic}. It may happen then that the packet would be first forwarded from Boston to Shanghai, and only then from Shanghai to Los Angeles. The packet would thus travel over a large geographic distance and, therefore, accumulate large time delay.

To address this suboptimality, we split the whole geographic area of interest into \emph{regions}. These regions are not to be confused with smaller \emph{zones} defined earlier. Within each region we then establish \emph{local hub} zones. These hub zones, from all the regions, are all mapped close to the origin of $\mathbb{H}^3$, so that, a majority of geographically local paths are GGR-routed through them. If a packet from the above example could stay within the North America region, the overall time delay would be significantly reduced indeed.

Given any collection of $M$ regions, e.g., the one from the next section, the GH scheme is regionalized as follows. \emph{First}, map all zones to their corresponding regions. \emph{Second}, within each region, sort the zones belonging to the region in the decreasing order of their centrality scores, and pick the first $N_{hubs}$ zones from each list, where $N_{hubs}$ is a scheme parameter. This way, an ordered list $L_{k} = \{z_{1,k}, z_{2,k}, \ldots, z_{N_{hubs},k}\}$ of zones $z_{i,k}$, $i=1,2,\ldots,N_{hubs}$, is obtained for each region $k = 1,2,\ldots,M$. Each entry in this list corresponds to the $1$st, $2$nd, \ldots, $N_{hubs}$-th most central zone in each region. \emph{Third}, the centrality rank $t_i$ of zones $z_{i,k}$ are set to $i$, i.e., $t_{z_{i,k}}=i$. For example, \emph{all} the 2nd most central zones in \emph{every} region get the \emph{same} centrality rank $t=2$. \emph{Fourth}, given thus computed centrality ranks of all hub zones, their radial coordinates are still computed using Eq.~\eqref{eq:centrality}. This way, \emph{all} the 2nd most central zones in \emph{every} region get the \emph{same} radial coordinate $r=\ln(\xi+2)$, while the angular coordinates are still determined by geographic coordinates. Therefore, regional hub zones in RGH  become equally important across different regions, so that GGR favors geographically local hubs, thus reducing delays. The \emph{remaining} zones $z$ that do not appear in any of the hub lists $L_k$ are then sorted in the decreasing order of their centrality scores, and their centrality ranks are set to $t_z=N_{hubs}+i$, where $i$ is $z$'s rank in this sorted list of remaining zones. Their radial coordinates are still set using Eq.~\eqref{eq:centrality}.

The RGH scheme distorts slightly the radial coordinate distribution $P(r)$ from the pure exponential distribution. However, only a small fraction of zones is assigned radial coordinates differently than in the \emph{GH} scheme. We show in the next section that even though the SR in RGH is slightly lower than in GH, the delays in RGH are noticeably lower than in GH, nearly as low as in the delay-optimal GEO scheme.

\section{GGR performance evaluation}
\label{sec:simulations}

Here we evaluate the performance of GGR in hypothetical world-wide overlay network deployments and in the existing NDN testbed, using the network design scheme from Section~\ref{sec:design}, and the three addressing schemes: GEO, GH, and RGH.

\textbf{Performance metrics.} The GGR performance metrics that  we evaluate are the \emph{success ratio (SR)} and the 50-th and 95-th percentiles of \emph{overlay and underlay delay stretches (ODS and UDS)}. These delay stretches measure how close routing paths are to minimal-delay paths in the network. Given a source-destination pair of nodes in an overlay network, let $d_1$ be the delay along the GGR path between these nodes, $d_2$ be the delay along the shortest-delay path between the same nodes, and $d_3$ be the delay between them in the underlay. The ODS and UDS are then defined as $ODS = \frac{d_1}{d_2}$ and $UDS=\frac{d_1}{d_3}$. The 95-th percentiles of delay stretches provide statistics on the worst delays in the considered schemes. An ideal combination of addressing and topology would not only maximize SR, but also minimize ODS and UDS.

\textbf{Delay estimation.} Since we cannot measure link delays in synthetic networks, we analyzed the delays of links in the NDN testbed~\cite{ndntestbedurl} and found that the correlation between the measured delays and the geographic great circle distances is given by
    \begin{equation}\label{eq:delay}
        delay(ms) \approx \frac{distance(km) + 1165}{49}
    \end{equation}
We use this relation to estimate the underlay delay between any pair of nodes, or equivalently the delay of a direct overlay link between the two nodes, if it exists. This relation is consistent with the more thorough measurements in~\cite{landa2013large}.

\textbf{Centrality scheme.} We assume that zones $z$ are cities in the world, and set the centrality score $s_z$ of city $z$ equal to the \emph{city population}. The most populated cities thus have the smallest radial coordinates. Other possibilities for centrality scores are discussed in Section~\ref{sec:discussion}.

\textbf{Regions in RGH.} We split the whole Earth surface into regions using administrative level~1 units and their populations---these units are states in the USA, or \emph{oblasts} in Russia, for instance. These units are then merged into larger contiguous regions of approximately equal populations. If every region has approximately the same population, then the total network load within each region would be approximately the same, since traffic volumes are positively correlated with geographic site populations~\cite{betker2014comprehensive}.

\textbf{Data.} The city data including geographic positions and populations are taken from the GeoNames database~\cite{Geonames}. Administrative level 1 units and their boundaries are taken from the free NaturalEarth database~\cite{natearth}. Gridded population data is obtained from SEDAC UN-adjusted population database~\cite{sedac2016}.

\textbf{Scheme parameters.} The following model parameters are used in the evaluation:
    \begin{compactitem}
        \item{\emph{All models}: the number of connections that the new nodes establish in the network design scheme is set to $m = 5$. This is the minimal value of $m$ that guarantees a nearly 100\% SR in the \emph{RGH} scheme.}
        \item{\emph{GH}: $\xi = 5$. This choice ensures that there are no ``superhubs'' attracting a finite fraction of all new connections due to their small radial coordinates given by~(\ref{eq:centrality}).}
        \item{\emph{RGH}: $\xi = 5$, $M = 14$ and $N_{hubs} = 5$. This choice of the number of regions $M$ allows to divide the Earth into regions with approximately $500$ millions people in each region. This division is optimal for the given number of local hubs.}
    \end{compactitem}

\textbf{Node appearance order.} To sensibly compare the three addressing schemes, we have to fix the node appearance order in all of them. We expect that in new network deployments, new nodes are more likely to appear first in more populated cities, where the user demand for network services is higher. This expectation is confirmed by real-world network data. For instance, the number of the Internet routers is positively correlated with the population size of the geographical locations~\cite{lakhina2003geographic}; the Internet traffic volume is also correlated with the population density of geolocations~\cite{betker2014comprehensive}. Therefore for simplicity we assume that the network grows with nodes appearing in cities, one node per city at a time, in the order given by the city centrality ranks, i.e., in the decreasing order of city population sizes.

To demonstrate stability of GGR's performance w.r.t.\ the node appearance order, we also consider a more realistic non-deterministic appearance order in the RGH scheme. Specifically, we place the local hub nodes as the \emph{RGH} prescribes, but then new nodes appear in cities with probabilities proportional to the city population. This process mimics more closely a real-world deployment scenario, where the appearance order of nodes is not deterministic, but the probability of node appearance is still proportional to the number of potential network users. We will see that the deterministic and non-deterministic node appearance orders yield very similar results.

\textbf{Networks.} Two types of networks are considered. The first one is the real operational NDN testbed~\cite{ndntestbedurl}, consisting of 33 NDN sites as of November 2016. We ignore its current topology, and rebuild it from scratch using the three considered addressing schemes. We use the same network design scheme from Section~\ref{sec:design} for each addressing scheme. Even though the design scheme is the same, the resulting topologies are different since the addressing schemes are different. The second type of networks are hypothetical world-wide overlay deployment networks that are generated using the same schemes applied to all possible cities present in the GeoNames database, $\sim 112,000$ cities total.

The main purpose of using these hypothetical networks is to evaluate GGR's \emph{scalability}, i.e., to tell how well GGR's performance metrics \emph{scale} with the growing network size. We grow these networks according to the network design and addressing schemes, following the node appearance order discussed above, and measure GGR's performance metrics at certain network sizes. The smallest network size when the measurements are performed is $205$, when all $N_{hubs} = 5$ local hubs from all the $M = 14$ regions in the RGH scheme are added to the network. The last measurements are performed when the networks are grown up to $10,000$ nodes.

\textbf{Connectivity failures.} In the NDN testbed, we evaluate GGR's performance degradation under all possible 1-node and 1-link failures, thanks to the testbed's small size. That is, one node or one link is first removed from the topology, emulating link or node failure, and all the GGR's performance metrics---SR, UDS, ODS---are then measured. We measure these metrics for \emph{all} possible 1-node and 1-link removals, and report the results averaged over all such trials. If 1-node/1-link removals do not strongly affect routing measurements even in this small topology, one cannot expect such small network damage to be noticeable in larger topologies. Second, much more catastrophic connectivity disruptions are investigated,  both in the NDN testbed and in hypothetical networks: $20\%$ of all links are removed at random, and all the GGR's performance metrics are then measured again. In hypothetical networks, these measurements are performed at different network sizes mentioned above. In the NDN testbed, we perform $100$ trials of $20\%$ random link removals, to sample a majority of failed network configurations. The results reported for the NDN testbed are averaged over these $100$ trials.

The main purpose of these tests is to check how much GGR suffers from connectivity failures, i.e., how many paths become non-routable and should thus be taken care of, if needed, by an auxiliary routing strategy. If routing SR stays close to 1, only a tiny fraction of paths needs to be corrected by such a strategy, meaning that routing control message overhead remains close to zero.

\begin{figure}[htp]
    \centering
    \includegraphics[width=\columnwidth]{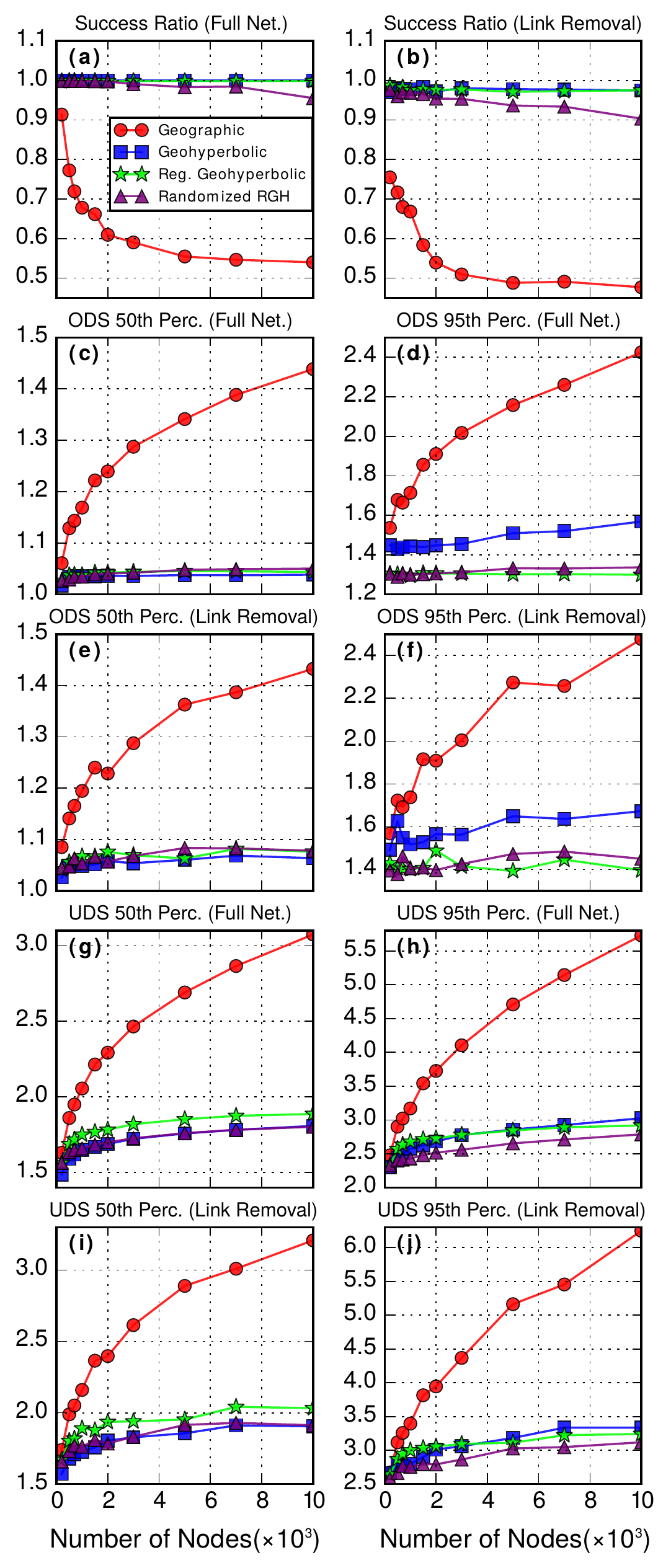}
    \caption{Evaluation of GGR in the GEO, GH, RGH schemes, and in the RGH scheme with randomized node arrivals, applied to hypothetical world-wide overlay networks. ``Full net.'' and ``Link removal'' in the figure titles corresponds respectively to the undamaged networks and the same networks with $20\%$ of all links randomly removed. The routing performance metrics are the Success Ratio (SR) and the 50-th and 95-th percentiles of Overlay and Underlay Delay Stretches (ODS and UDS). All the metrics are measured as functions of the network size shown on the horizontal axes.}
    \label{fig:plots}
\end{figure}

\textbf{Evaluation results.} The results of these experiments in hypothetical overlay networks are reported in Fig.~\ref{fig:plots}. Both GH and RGH show excellent  routing overhead scaling, because their success ratios are very close to $1$, and nearly independent of the network size. The median ODS delay stretch does not noticeably depend on the network size either, quickly settling to a bounding value around $1.05$. In contrast, the GEO scheme degrades fairly quickly, thus becoming unfeasible to use in real networks. The networks built according to the GEO scheme are thus much more fragile, compared to the (R)GH schemes, w.r.t.\ connectivity disruptions, due to the lack of the notion of node centrality, and consequently the lack of hyperbolicity.

Both the ODS 95-th percentiles show that the RGH scheme does noticeably reduce, compared to GH, the delays of the worst-delay-stretch paths. The results of the more realistic scenario with the randomized node appearance order used in the RGH scheme are comparable to those using the deterministic node appearance order. The ODS $95$-th percentiles appear to be bounded by a value around $1.3$ as networks grow according to RGH, with or without randomized node appearance. Most importantly, even the catastrophic connectivity disruptions do not much affect the success ratio and delay stretches of the (R)GH schemes, thus making them feasible schemes for the deployment in real networks.

The results for the NDN testbed are presented in Table~\ref{tab:ndntab}. The GH and RGH's SR is exactly $1$ in the original NDN testbed, as well as in \emph{all} possible 1-node- or 1-link-failed testbeds, meaning that no auxiliary routing protocol is needed to handle such failures, and the routing message overhead is \emph{exactly zero}. Even if $20\%$ of links fail, both schemes maintain success ratio remarkably close to optimal~$1$ (0.97-0.98). We emphasize that these results imply that even in highly dynamic scenarios with catastrophic link failures, only a small fraction of paths become GGR-unroutable, and must thus be handled by alternative routing methods if needed. In normal conditions, or when only one node or link fails, such alternative methods are \emph{not} required and the routing overhead is \emph{strictly zero}, because all paths remain to be GGR-routable. That is, GGR finds alternative paths around failures \emph{without} any recomputations or information exchange between routers. The median overlay stretch is $1$ in both the original and failed networks, meaning that at least a half of GGR's paths are delay-optimal even upon major changes in network topology, while $95\%$ of paths experience the ODS below $2.6$ in GH and $1.5$ in RGH in the undamaged topology, compared to GEO's $1.3$. These numbers show that, as in synthetic networks, RGH helps to decrease GH's delay to almost the GEO values, without sacrificing $100\%$ SR. In damaged topologies, the ODS $95$-th percentiles are not significantly larger than in the undamaged one. The UDS is always larger than the ODS as expected, but not by much for the majority of paths---the undamaged median, for instance, is $1.2$ for all the GEO, GH, and RGH, while overall the worst numbers are the UDS $95$-th percentiles in the catastrophically damaged topologies, which are $2.0$, $4.5$, and $3.8$ for GEO, GH, and RGH, respectively.

\textbf{Code and data availability.} All the code and data used in the experiments described above are freely available at~\cite{code}.

\begin{table}[htp]
\centering
\caption{Evaluation of GGR in the GEO, GH, and RGH schemes, applied to the NDN testbed. The SR, ODS, and UDS metrics shown for damaged networks are averaged over all possible 1-link or 1-node removals from the original NDN testbed topology, and over $100$ trials of removal of $20\%$ random links.}
\resizebox{0.48\textwidth}{!}{
\begin{tabular}{|c|c|c|c|c|c|c|}
\hline
\rotatebox[origin=c]{90}{ Model }                     & Scenario     &   SR    & ODS 50th & ODS 95th & UDS 50th & UDS 95th  \\ \hhline{|=|=|=|=|=|=|=|}
\multirow{4}{*}{\rotatebox[origin=c]{90}{GEO}}      & Original     &  0.895  &  1.000  &  1.259  &  1.166  &  1.675  \\ \cline{2-7}
                                                    & 1 node       &  0.893  &  1.000  &  1.263  &  1.166  &  1.682  \\ \cline{2-7}
                                                    & 1 link       &  0.895  &  1.000  &  1.263  &  1.167  &  1.682  \\ \cline{2-7}
                                                    & $20\%$ links &  0.879  &  1.000  &  1.392  &  1.248  &  1.974  \\ \hhline{|=|=|=|=|=|=|=|}
\multirow{4}{*}{\rotatebox[origin=c]{90}{GH}}       & Original     &  1.000  &  1.000  &  2.646  &  1.225  &  3.616  \\ \cline{2-7}
                                                    & 1 node       &  1.000  &  1.000  &  2.655  &  1.230  &  3.725  \\ \cline{2-7}
                                                    & 1 link       &  1.000  &  1.000  &  2.647  &  1.229  &  3.660  \\ \cline{2-7}
                                                    & $20\%$ links &  0.977  &  1.000  &  2.931  &  1.309  &  4.511  \\ \hhline{|=|=|=|=|=|=|=|}
\multirow{4}{*}{\rotatebox[origin=c]{90}{RGH}}      & Original     &  1.000  &  1.000  &  1.544  &  1.216  &  2.187  \\ \cline{2-7}
                                                    & 1 node       &  1.000  &  1.000  &  1.607  &  1.219  &  2.337  \\ \cline{2-7}
                                                    & 1 link       &  1.000  &  1.000  &  1.578  &  1.219  &  2.209  \\ \cline{2-7}
                                                    & $20\%$ links &  0.970  &  1.000  &  2.161  &  1.268  &  3.826  \\ \hline
\end{tabular}
}
\label{tab:ndntab}
\end{table}
\vspace{-5mm}

\section{Discussion}
\label{sec:discussion}

We emphasize that the considered schemes are not directly applicable to existing underlays, such as the Internet, simply because the key element of these schemes is the network design scheme that prescribes how connections must be formed in the network, given the geohyperbolic addresses of nodes/routers in it. This lack of freedom in engineering the topology arbitrarily is the price paid for exceptional routing scalability. Nevertheless the specified network design scheme does grow Internet-like topologies, and even stronger, the real Internet topology follows, albeit approximately, the specified network design scheme, as shown in~\cite{boguna2010sustaining,Papadopoulos2012}, as if the actual evolution of the real Internet has followed this design scheme at a large scale but with some random deviations at smaller scales.

The proposed schemes can be used in new underlays however, in which geographic positions of nodes are known, and which can follow the specified network design scheme. For initial deployment in those cases, one has first to fix a centrality score scheme, such as population density. The only requirement to the centrality score of a geolocation is that it should be uniquely determined by the geolocation. If so, then any node at any geolocation $(\theta,\phi)$ has a unique location $(r,\theta,\phi)$ in the geohyperbolic space. The network design scheme prescribes that if a new node at a given geolocation is to be added to the network, then this node \emph{must} be linked to a fixed number $m$ of already existing nodes in the network that are closest to the new node in the geohyperbolic space. To determined this set of $m$ nodes at initial stages of deployment, the geolocation of all existing nodes in network must probably be globally known. At later stages of deployment however, when the network grows large, the geohyperbolic space can be tessellated into small cells, so that the determination of the set of $m$ existing nodes to which the new node must connect, can be limited to within the cell in which new node appears, e.g.\ similar to how this is done in fast generation of hyperbolic graphs~\cite{VonLooz2016,bringmann2015sampling}. If any node fails, it is simply removed from the network along with all its connections. If the node comes back, its connections are restored as usual. As the results presented above demonstrate, in a vast majority of cases this transient network dynamics will not incur \emph{any} routing overhead or stretch increase.

New nodes in the network can certainly appear not necessarily within a city, but at any geolocation. In this case, the zones $z$ required for the calculation of centrality scores $s_z$ can be obtained using Voronoi tessellation with cities as base points, for instance.

We used city population sizes as a centrality measure for cities/zones, but other measures can be chosen for different network deployments. For instance, it may happen in some networks that the traffic load is driven not by many small users, but by few large users that use network resources extensively. In this case the centrality scores can be based on expected users' activity averaged over large time scales.

Dynamics of centrality scores in the (G)HR schemes can affect the efficiency of GGR. However, the topology dynamics time scale is likely to be much smaller than the time scale of centrality dynamics in most cases, so that centrality score recalculations will likely be rarely needed. For instance, the dynamics of city populations occurs on the  scale of decades with a slowing rate~\cite{abundo2013city}.

The tiny fraction of unsuccessful paths in GH and RGH schemes in failing networks can be handled by auxiliary routing strategies, such as adaptive forwarding protocols~\cite{lehman2016experimental}. As shown in~\cite{lehman2016experimental}, if the SR is maintained close to $100\%$, these protocols do not introduce any noticeable protocol message overhead. Alternatively, parameter $m$ can simply be increased, resulting in denser and more robust network topologies.

\section{Related Work}
\label{sec:related}

The literature on greedy geometric routing in general, and geographic routing in particular, and on overlay virtual coordinate systems and network designs, is extremely vast, so that here for brevity we mention only the most pertinent results, other than those already referred to. The fact that any given network topology can be mapped to the hyperbolic plane such that GGR's SR is $100\%$ was first proven in~\cite{kleinberg2007geographic}. The proof goes via building a spanning tree in the network, and then isometrically embedding the tree into the hyperbolic plane such that SR is $100\%$. Unfortunately, as soon as any link belonging to this tree fails, the tree must be recomputed from scratch. If such recomputations are to be performed in a distributed manner, they would incur significant overhead. Hyperbolic network mapping algorithms that do not require global recomputations upon network topology changes were introduced in~\cite{cvetkovski2009hyperbolic}. The key idea behind those algorithms is to rely on gravity-pressure routing to route around failing links. Multiple extensions of dynamic overlay mapping to hyperbolic spaces were investigated in~\cite{cassagnes2011overlay,magoni2015dynamic}, where, as in all the previous work, the hyperbolic spaces are \emph{virtual}, unrelated to geography. In a recent outstanding theoretical contribution~\cite{bringmann2016greedy}, it was \emph{proven} that both the SR and stretch of GGR and similar algorithms in random hyperbolic graphs converge, in the $n\to\infty$ limit, to optimal~$1$ in a wide range of parameters. Unfortunately, the constants, as functions of model parameters, characterizing the speed of this convergence remain unknown. Therefore as far as finite-size networks are concerned, only experimental results, such as those presented in this paper, appear to be currently available.

\section{Conclusion}
\label{sec:conclusion}

We introduced simple geohyperbolic addressing schemes, and a simple network design scheme coupled to them. The schemes do not require any sophisticated non-local computations of addresses of nodes joining the network. The addresses are determined only by geographic coordinates of nodes, and by centrality scores defined by these coordinates. Greedy geometric routing in resulting network topologies is remarkably scalable and robust with respect to network growth and failures. The FIB sizes are as small as they can be, assuming some per-adjacency information must be stored anyway, while the routing protocol message overhead is nearly zero. The delay of greedy routing paths is also very close to optimal.

The price paid for this remarkable routing scalability, robustness, and efficiency is that the network topology cannot be arbitrary but must follow the network design scheme. If extreme routing scalability is a requirement, e.g., in IoT-networks, then the proposed schemes are viable candidates for deployment in real-world networks, where node geolocation information is available, and where network topology construction can follow the simple specified design rules.

\textbf{Acknowledgments.} We thank V.~Jacobson, L.~Zhang, B.~Zhang and D.~Oran for useful discussions and suggestions. This work was supported by NSF CNS-1345286, CNS-1344495, and CNS-1442999.

\balance
\bibliography{bib}
\bibliographystyle{abbrvurl}

\end{document}